\documentclass[twocolumn,english,aps,prl,superscriptaddress,floats,nobibnotes]{revtex4-1}
\usepackage[latin9]{inputenc}
\usepackage{color}
\usepackage{babel}
\usepackage{bm}
\usepackage{bbm}
\usepackage{amsmath}
\usepackage{amssymb}
\usepackage{graphicx}
\PassOptionsToPackage{version=3}{mhchem}
\usepackage{mhchem}
\usepackage{braket}
\usepackage[unicode=true,
 bookmarks=false,
 breaklinks=false,pdfborder={0 0 1},backref=false,colorlinks=false]
 {hyperref}
\hypersetup{
 colorlinks,linkcolor=blue,citecolor=blue,urlcolor=blue}

\makeatletter

\usepackage{babel}
\PassOptionsToPackage{version=3}{mhchem}

\usepackage{newfloat}
\usepackage{latexsym}
\usepackage{dcolumn}
\usepackage{float}

\usepackage{bm}
\usepackage{babel}

\makeatother

\begin{document}

\title{Gauge phonon dominated resistivity in twisted bilayer graphene near magic angle}

\author{Indra Yudhistira}
\thanks{These two authors contributed equally to this work}
\affiliation{Centre for Advanced 2D Materials, National
University of Singapore, 6 Science Drive 2, 117546, Singapore}
\affiliation{Department of Physics, National
University of Singapore, 2 Science Drive 3, 117551, Singapore}

\author{Nilotpal Chakraborty}
\thanks{These two authors contributed equally to this work}
\affiliation{Yale-NUS College, 16 College Avenue West, 138527, Singapore}

\author{Girish Sharma}
\affiliation{Centre for Advanced 2D Materials, National
University of Singapore, 6 Science Drive 2, 117546, Singapore}
\affiliation{Department of Physics, National
University of Singapore, 2 Science Drive 3, 117551, Singapore}

\author{Derek Y. H. Ho}
\affiliation{Centre for Advanced 2D Materials, National
University of Singapore, 6 Science Drive 2, 117546, Singapore}
\affiliation{Yale-NUS College, 16 College Avenue West, 138527, Singapore}

\author{Evan Laksono}
\affiliation{Centre for Advanced 2D Materials, National
University of Singapore, 6 Science Drive 2, 117546, Singapore}

\author{Oleg P. Sushkov}
\affiliation{School of Physics, The University of New South Wales, Sydney 2052, Australia}

\author{Giovanni Vignale}
\affiliation{Centre for Advanced 2D Materials, National
University of Singapore, 6 Science Drive 2, 117546, Singapore}
\affiliation{Yale-NUS College, 16 College Avenue West, 138527, Singapore}
\affiliation{Department of Physics and Astronomy, University of Missouri, Columbia, Missouri 65211, USA}

\author{Shaffique Adam}
\email{shaffique.adam@yale-nus.edu.sg}
\affiliation{Centre for Advanced 2D Materials, National
University of Singapore, 6 Science Drive 2, 117546, Singapore}
\affiliation{Department of Physics, National
University of Singapore, 2 Science Drive 3, 117551, Singapore}
\affiliation{Yale-NUS College, 16 College Avenue West, 138527, Singapore}

\date{\today}
\begin{abstract}
Recent experiments on twisted bilayer graphene (tBG) close to magic angle show that a small relative rotation in a van der Waals heterostructure greatly alters its electronic properties.  We consider various scattering mechanisms and show that the carrier transport in tBG is dominated by a combination of charged impurities and acoustic gauge phonons.  Charged impurities still dominate at low temperature and densities because of the inability of Dirac fermions to screen long-range Coulomb potentials at charge neutrality; however, the gauge phonons dominate for most of the experimental regime because although they couple to current, they do not induce charge and are therefore unscreened by the large density of states close to magic angle.  We show that the resistivity has a strong monotonically decreasing carrier density dependence at low temperature due to charged impurity scattering, and weak density dependence at high temperature due to gauge phonons.  Away from charge neutrality, the resistivity increases with temperature, while it does the opposite close to the Dirac point.  A non-monotonic temperature dependence observed only at low temperature and carrier density is a signature of our theory that can be tested in experimentally available samples.    
\end{abstract}

\maketitle

The remarkable observations of superconductivity and insulating behaviour near magic angle in twisted bilayer graphene (tBG)~\cite{cao_unconventional_2018,cao_correlated_2018,yankowitz_2019_tuning} have underlined the importance of twist angle as an additional control knob in van der Waals heterostructures. The band structure of tBG can be significantly altered with just small variations in the twist angle \cite{laksono2018singlet}.  Special cases emerge near the so called ``magic angles" when the lowest energy bands become almost flat~\cite{bistritzer2011moire}, providing a platform for exotic physics arising from strong correlations.  These recent experiments have inspired a large body of theoretical work which aim to understand both the origin of this strongly correlated phase~\cite{xu_topological_2018,roy_unconventional_2018,po_origin_2018,koshino_maximally_2018,kang_symmetry_2018,padhi_doped_2018,guo_pairing_2018,liu_chiral_2018,isobe_unconventional_2018,you_superconductivity_2018,gonzalez_kohn-luttinger-2019,xie_on_2018} and phonon-driven superconductivity in tBG~\cite{lian_twisted_2018,wu_theory_2018,choi_strong_2018}. More recent experimental work  has focused on electron transport~\cite{cao_strange_2019,sharpe_emergent_2019}.  A striking characteristic of transport in tBG close to magic angle is its extremely high and $T$-linear resistivity at high temperatures~\cite{cao_unconventional_2018, yankowitz_2019_tuning}. Understanding the physics of these transport features could provide the insights necessary to understand the observed strongly correlated phases. 

Carrier transport in monolayer and bilayer graphene has been studied extensively over the past decade, both in theory and in experiment~\cite{sarma2011electronic}. Traditionally, charged impurities dominate the electronic carrier transport at low temperature for both graphene monolayers and bilayers, and these also induce fluctuations in the carrier density close to the Dirac point. Acoustic phonons become relevant at intermediate temperatures ($T \gtrsim 100~\rm{K}$) and give rise to a characteristic linear in $T$ resistivity.  At still higher temperatures ($T \gtrsim 250~\rm{K}$) optical phonons take over as the dominant scattering mechanisms.  While the linear-in-T resistivity in tBG has been previously attributed to phonons \cite{wu2018phonon}, the magnitude of the deformation potential extracted from experiment and the relevant temperature scales do not match what we know from our extensive studies on graphene~\cite{sarma2011electronic}.    

In this Letter, we present a complete theory for transport in tBG at low densities near magic angle tBG ($\theta_M$), identifying all the relevant scattering mechanisms responsible at various temperatures probed experimentally.  We find that for tBG close to magic angle, there is a crossover from charged-impurity limited transport to phonon-limited transport.  Usually, phonon-dominated resistivity is modeled by the deformation potential. However, we show that the deformation potential contribution becomes irrelevant due to screening, and instead the phonon contributions arise from a gauge-field term~\cite{sohier_phonon-limited_2014}.  The dominance of these gauge phonons arise due to the immunity of these particular phonons to the enhanced screening from the flat bands in tBG.  

The enhanced screening reduces the importance of all other scattering mechanisms including that of charged impurities and deformation potential phonons.  Our theory shows that away from charge neutrality, the resistivity {\it increases} linearly with temperature with weak dependence on both carrier density and impurity density (and therefore shows little sample-to-sample variations).  However, close to the Dirac point at sufficiently low temperature, a non-monotonic in temperature and strong density dependence reveals the role of charged impurities. 

\begin{figure}[t]
	\includegraphics[scale=0.47]{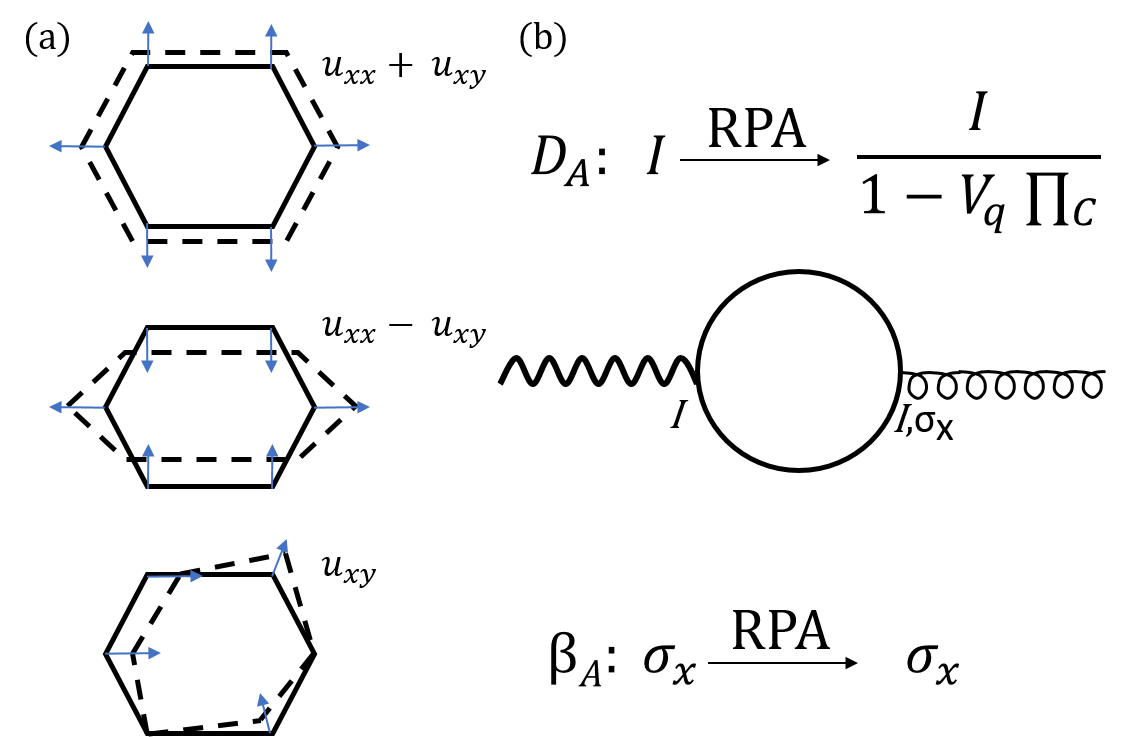}
	\caption{Scalar and gauge deformations of twisted bilayer graphene (a) and how they are screened within the random phase approximation (b).  The uniform stretching of the moir\'e Brillouin zone (top sketch) couples to both charge and current and is strongly screened by the electrons in the flat moir\'e bands at low energy.  The asymmetric and shear modes (bottom two sketches) are area preserving and therefore act like a gauge potential that couple to current but not charge.  As a result, these remain unscreened even as the Fermi velocity vanishes.  Wavy, solid and springy lines refer to Coulomb interaction $V_q$, electrons and phonons, respectively.  $\Pi_C$ is the RPA polarization bubble. }
		\label{Fig:Feyn}
\end{figure}

As shown in Fig.~\ref{Fig:Feyn}, acoustic phonons in graphene can be classified as either deformation potential modes $D_A$ that change the area of the Brillouin zone, or gauge field modes $\beta_A$ that do not.  The $D_A$ contribution comes from the longitudinal acoustic phonons and causes charge separation that is susceptible to electronic screening. However, we show below that the gauge field term, which comes from both the longitudinal and transverse phonons, is unaffected by screening because although they couple to the current, they do not couple to charge.  These modes remain unscreened even as the Fermi velocity vanishes.  Charged impurity scattering that has dominated the transport properties of 2D materials are screened by the enhanced density of states.  As explained below, this weaker electron-impurity interaction is compensated by a lower velocity (see Table~\ref{table:1}), so in the end, the qualitative behavior of electron-impurity resistivity is the same as in monolayer graphene.  Due to the vanishing density of states of Dirac fermions as $n \rightarrow 0$ and $T \rightarrow 0$, charged impurity scattering still dominates the transport properties for sufficiently low carrier density and temperature.         

\begin{table}[]
\caption{Relevant scattering mechanisms and the dependence of resistivity on Fermi velocity $\rho(v^*)$.  Only the $\beta_A$ gauge phonons grow in importance as $v^* \rightarrow 0$ when $\theta \rightarrow \theta_M$.} 
\centering{}%
\begin{ruledtabular}
\begin{tabular}{cccc} 
Mechanism & Unscreened & Screening $\epsilon(q)^{-2}$ & Net effect \tabularnewline
\hline\tabularnewline
Charged imp. & $v^{*-2}$ & $v^{*2}$  & constant\tabularnewline
Phonons ($D_A$) & $v^{*-2}$ & $v^{*2}$  & constant\tabularnewline
Phonons ($\beta_A$) & $v^{*-2}$ & No screening  & $1/v^{*2}$\tabularnewline
Viscous friction & $v^{*-2}$ & $v^{*2}$  & constant\tabularnewline
Puddles & $v^{*-2}$ & $v^{*2}$  & constant\tabularnewline
\end{tabular}
\end{ruledtabular}
\label{table:1}
\end{table}

To qualitatively illustrate the role of screening in tBG, we consider
in Table~\ref{table:1} the Thomas-Fermi screening model where the polarizability is given by the density of states $N_D$.  Within a Dirac model (justified below), as $\theta \rightarrow \theta_M$, the linear bands become flat, $v^* \rightarrow 0$ and the static dielectric function diverges as $\epsilon(q) \sim  v^{*-1}$.  For most mechanisms, screening gives a $v^{*2}$ contribution, that is compensated by the usual $v^{*-2}$ dependence of resistivity on carrier velocity.  The inability of Dirac fermions to screen the gauge phonons dramatically increases their importance for the transport close to magic angle.  In particular, the crossover temperature for which gauge phonons dominate over charged impurities drops from $T_{\rm cross} \sim500~{\rm K}$ for monolayer graphene to $T_{\rm cross}~\sim5~{\rm K}$ for tBG.  It is remarkable that phonons which were traditionally neglected all the way until room temperature, now become important at such low temperatures.

The electron-phonon (e-ph) interaction for monolayer graphene within a single-valley Dirac model contains both scalar and vector potential components~\cite{sohier_phonon-limited_2014} and is given by $V_{e-ph} = \Phi I +  \boldsymbol{\sigma} \cdot  \mathbf{A}$, where 
\begin{eqnarray}
    \phi &=& D_A~(u_{xx} + u_{yy}) \\
    \mathbf{A} &=& \beta_A~ \left(\begin{array}{c} 
    u_{xx} - u_{yy} \\
    - 2u_{xy}
    \end{array} \right)
\end{eqnarray}

\begin{figure*}[hbt!]
\begin{centering}
\includegraphics[width=.66\columnwidth]{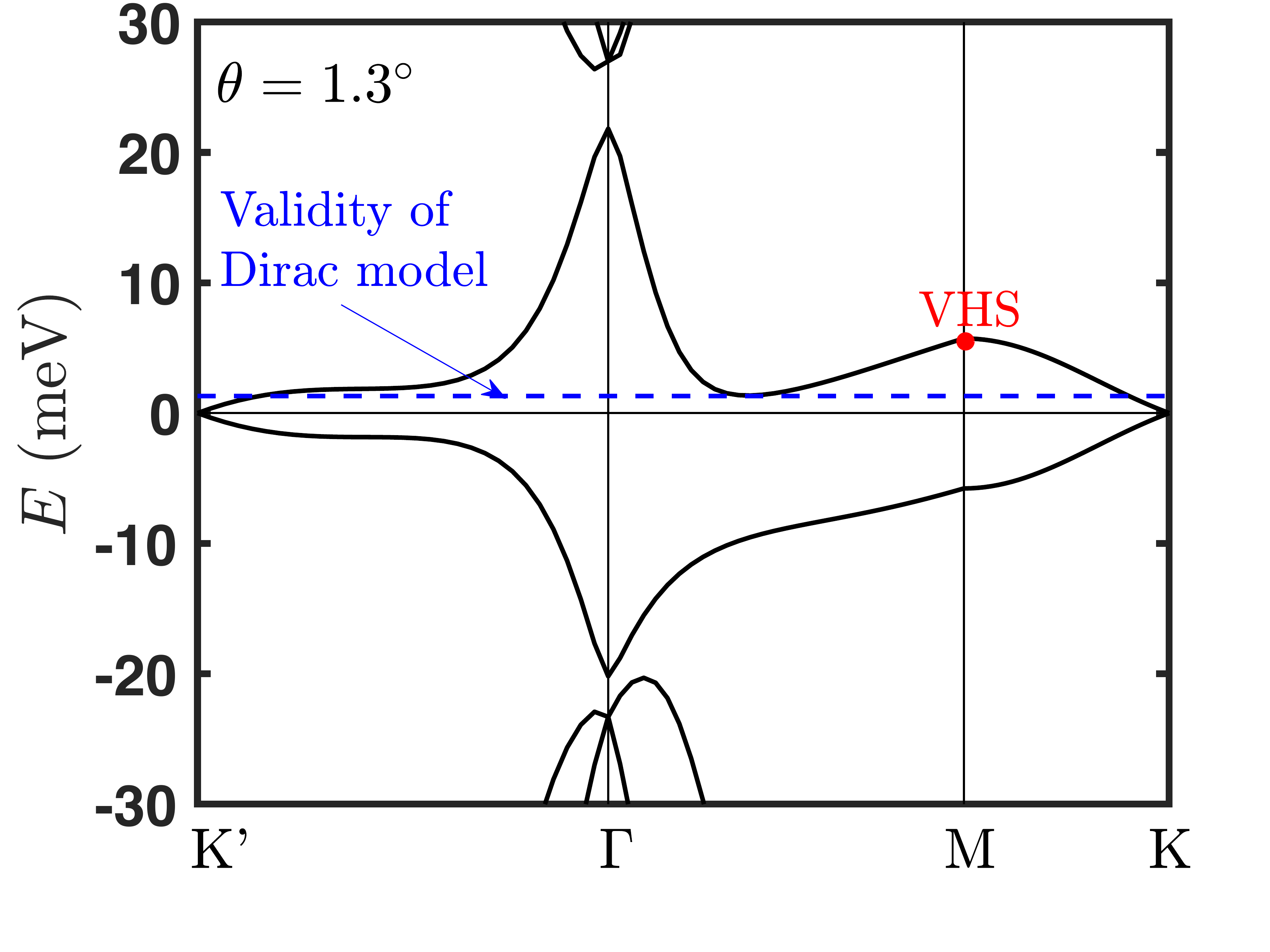}
\llap{\parbox[b]{4.5in}{\large{(a)}\\\rule{0ex}{1.4in}}}
\includegraphics[width=.7\columnwidth, height = 1.66in]{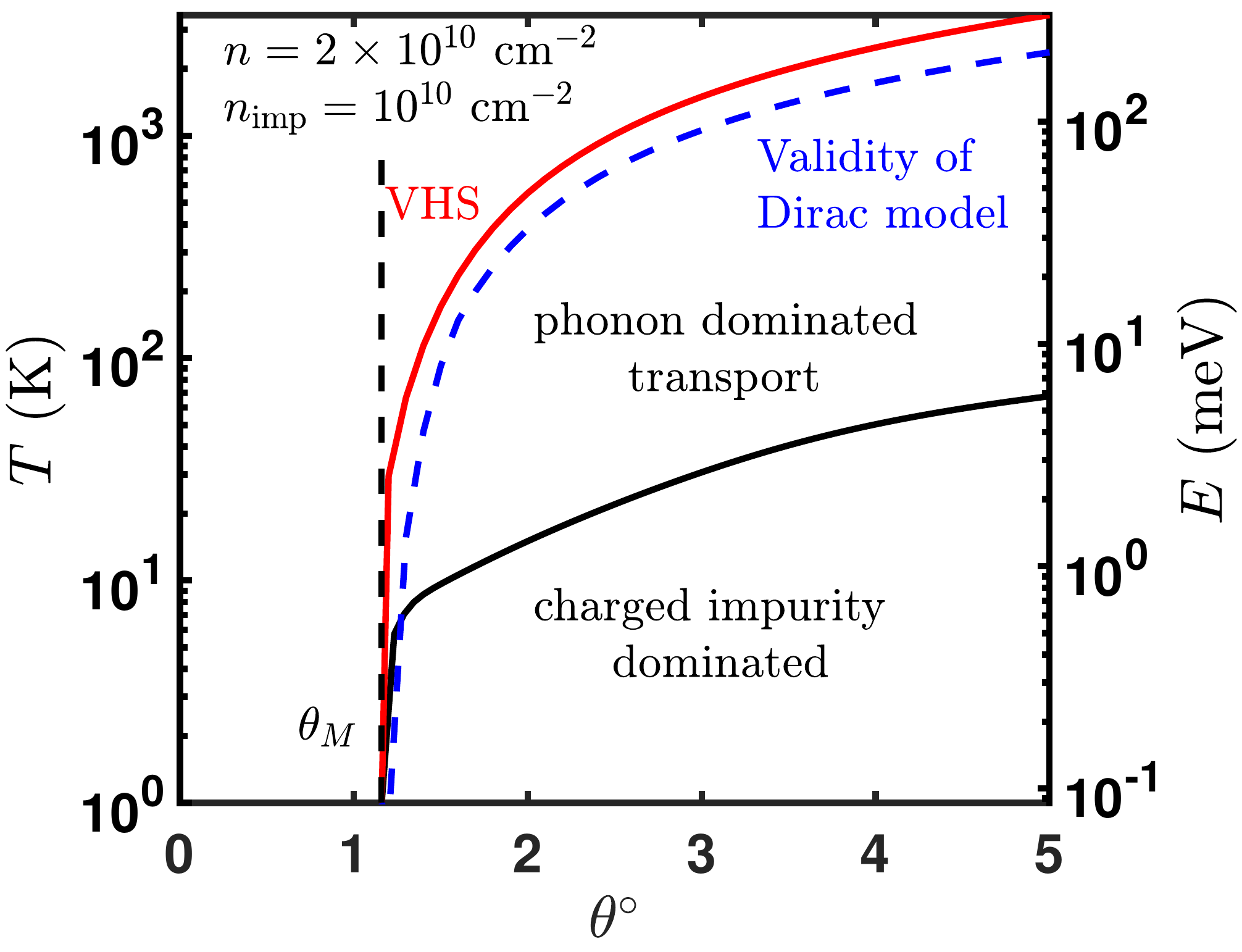}
\llap{\parbox[b]{4.8in}{\large{(b)}\\\rule{0ex}{1.4in}}}
\includegraphics[width=.63\columnwidth, height = 1.66in]{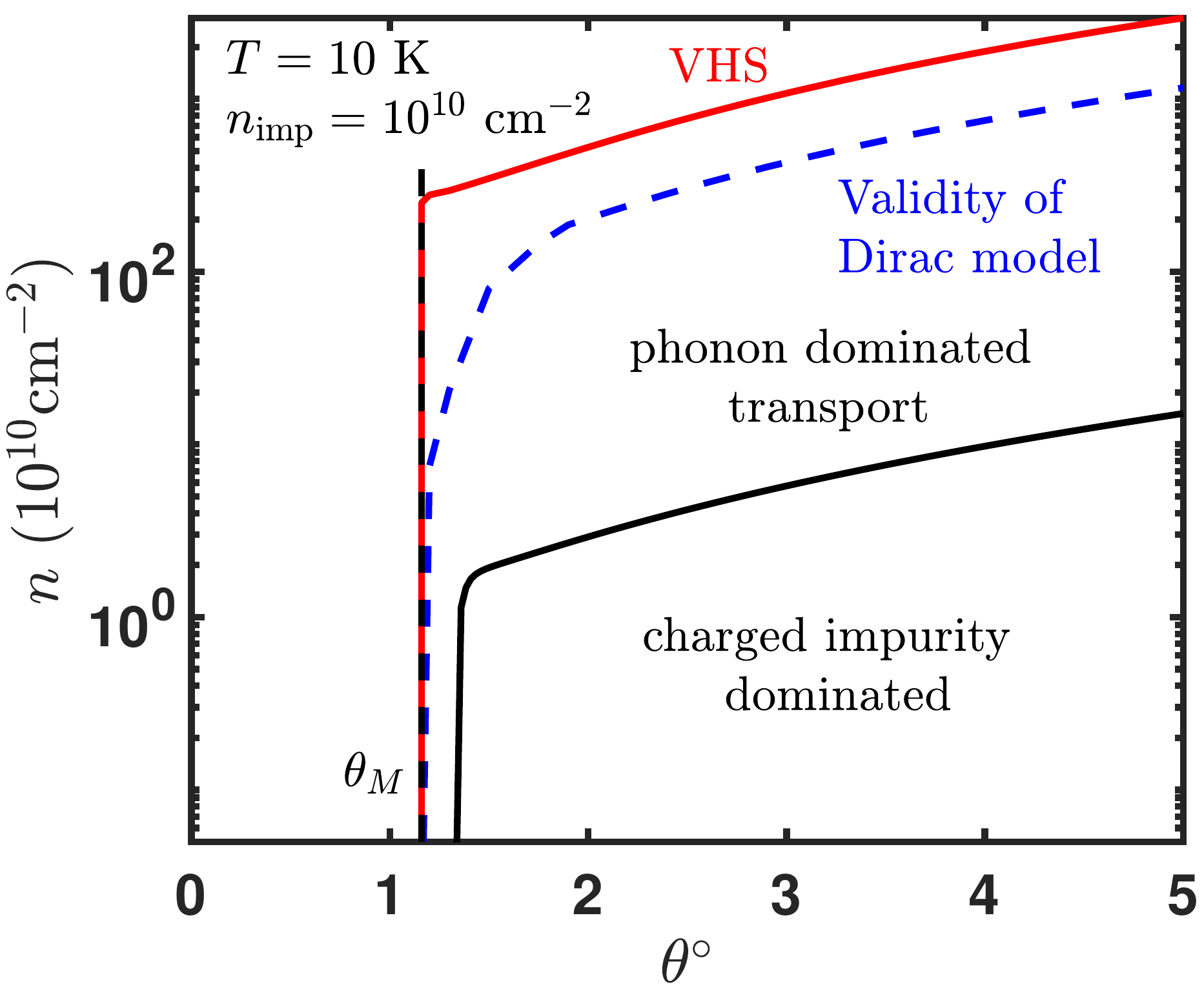}
\llap{\parbox[b]{4.3in}{\large{(c)}\\\rule{0ex}{1.4in}}}
\par\end{centering}
\caption{(Color online) (a) Electronic structure of twisted bilayer graphene.  A low energy effective Dirac Hamiltonian is valid for energies up to the blue line (marked Dirac) which for $\theta = 1.3^\circ$ is $\sim25$ percent of the van Hove singularity energy (marked VHS).  (b)  Gauge phonons dominate the transport at temperatures higher than $T_{\rm cross}$, while charged impurities dominate at lower temperatures.  This crossover occurs within the window probed in recent experiments.  (c)  Similarly, phonons dominate at high density, while charged impurities dominate at low carrier density.}
\label{diracvalid}
\end{figure*} 

\noindent where $D_A$ and $\beta_A$ are bare (unscreened) coupling constants. The electron-phonon vertex therefore has a two-dimensional matrix structure, and its diagonal (off-diagonal) part comes from the scalar (vector) potential.  While the diagonal component constitutes a scalar deformation potential, the off-diagonal component corresponds to distortions which do not induce any variation in the unit cell area, but rather induce bond length modulations which are often represented by a synthetic gauge field~\cite{oppen_gauge_2009}.  The random phase approximation (RPA) screening of the e-ph vertex by Coulomb interactions can be calculated from the polarizability bubble shown in  Fig.~\ref{Fig:Feyn}(b).  After including higher-order bubbles, the screening of a general vertex $g$ is given by the Dyson equation $g\rightarrow g + I V_q \Pi_g + I V_q \Pi_C V_q \Pi_g + \cdots$, where $I$ is the identity matrix, $V_q$ is the Fourier transform of the Coulomb interaction, $\Pi_C$ is the polarizability bubble, while $\Pi_g$ is the polarizability bubble with the vertex $g$ at the right end as shown in Fig.~\ref{Fig:Feyn}(b).  The RPA series is summed giving the screened vertex as 
\begin{equation}
    \beta_A \sigma_x \stackrel{\rm RPA}{\longrightarrow} \beta_A \sigma_x + \frac{\beta_A V_q \Pi_g}{1-V_q \Pi_C} \mathbbm{1}_2.
\end{equation}
\noindent The key result is that the off-diagonal component is unaffected by RPA screening i.e. the gauge phonons remain unscreened by the large density of states (these phonon modes could also be germane to the observed superconductivity).  We neglect the scalar component both because it is screened by $\Pi_C$ (similar to the $D_A$ phonons), and because $\Pi_g$ vanishes at charged neutrality.

To test the validity of the Dirac model, in Fig.~\ref{diracvalid} we plot the lowest energy bands for the continuum model~\cite{laksono2018singlet} at $\theta = 1.3^\circ$. The linear regime breaks down due to the emergence of Fermi pockets close to the $\Gamma$ point.  To stay within the regime of validity of the Dirac model, we limit our considerations to carrier densities $\lesssim 8 \times 10^{10}$ cm$^{-2}$ (although we expect our results to hold qualitatively for even higher densities).  In Fig.~\ref{diracvalid} we show that the crossover from charged impurity limited scattering to gauge phonon limited scattering occurs well within the regime where the Dirac Hamiltonian is valid.

\begin{figure*}[t!]
\begin{centering}
\includegraphics[width=\columnwidth]{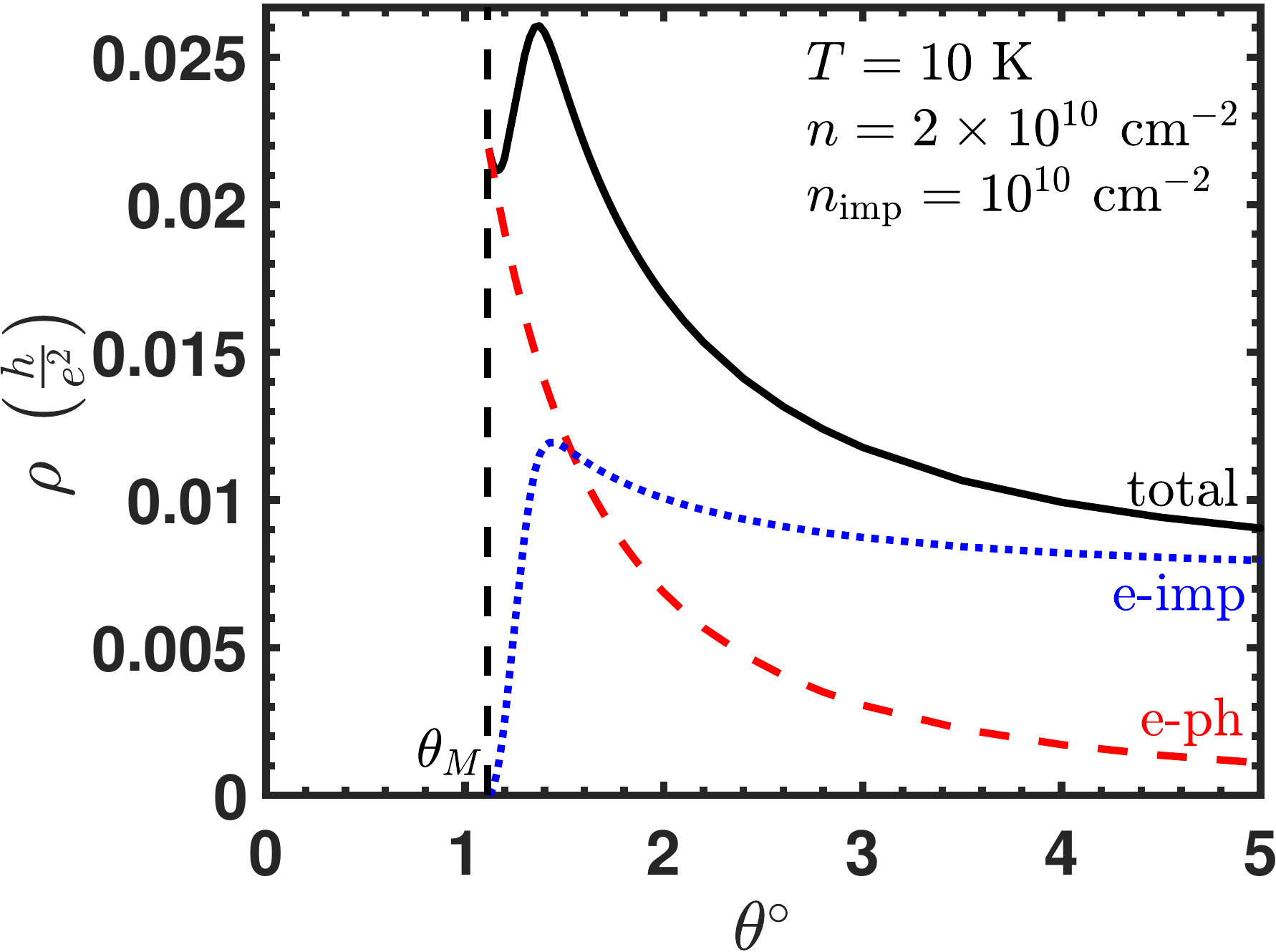}
\llap{\parbox[b]{6.8in}{\large{(a)}\\\rule{0ex}{2.35in}}}
\includegraphics[width=\columnwidth]{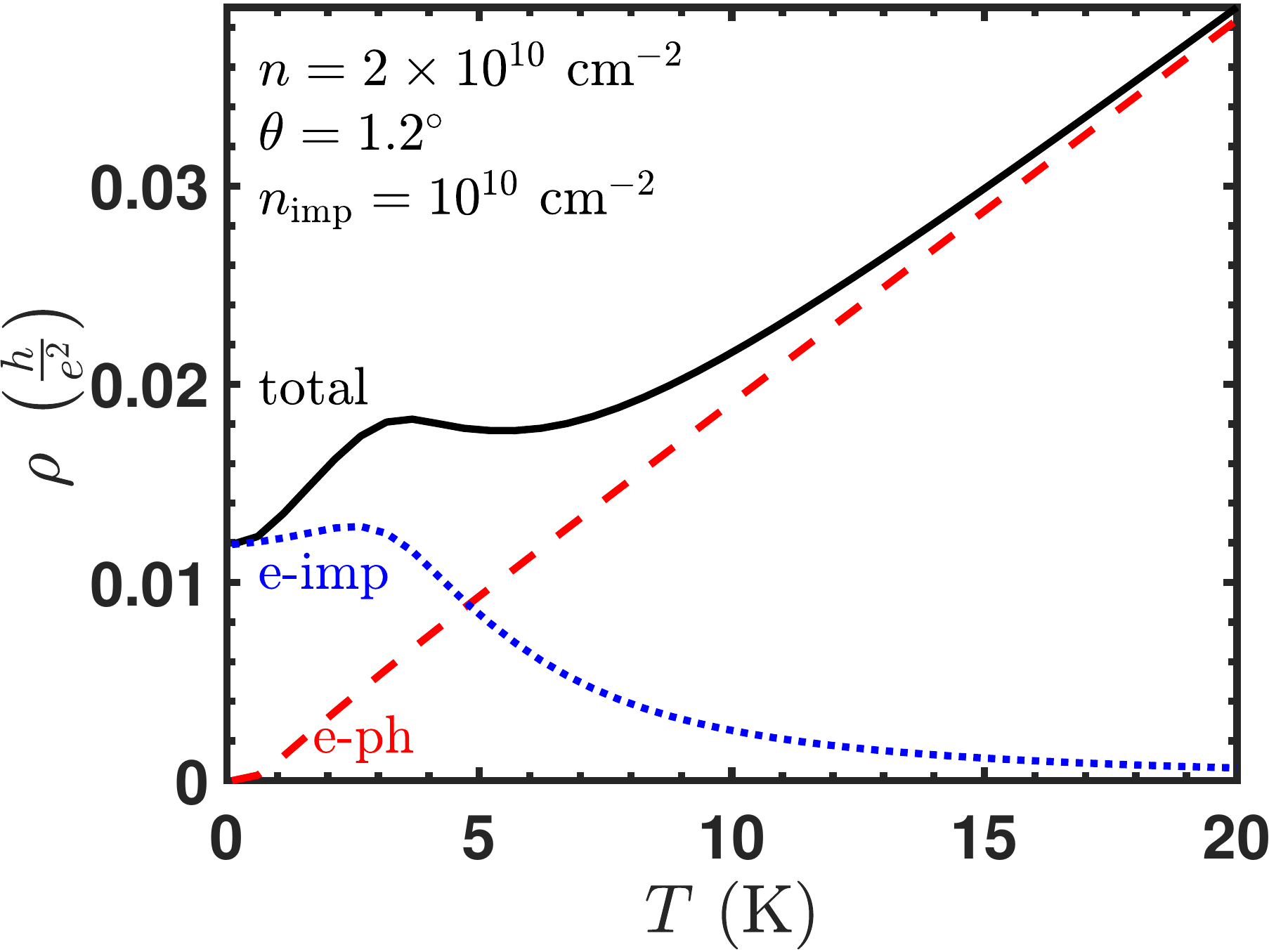}
\llap{\parbox[b]{6.6in}{\large{(b)}\\\rule{0ex}{2.35in}}}
\par\end{centering}
\caption{(Color online)  (a) Total resistivity as a function of twist angle (above magic angle), showing a crossover from phonon to impurity-dominated carrier transport. The black solid, red dashed and blue dashed lines are total, phonon and impurity limited resistivity, respectively. Gauge phonons are not screened by the large density of states associated with the flat bands, and dominate close to magic angle.  (b) Total resistivity as a function of temperature at fixed density and twist angle.  The crossover temperature in monolayer graphene is $\sim 500 K$, while near magic angle, the crossover temperature reduces to $5$~K due to the strong screening of charged impurities by the flat bands, and the immunity of gauge phonons towards screening.  The non-monotonicity arises from charged impurity scattering component crossing over from the degenerate ($k_{\rm B} T < \varepsilon_F$) to non-degenerate regime $k_{\rm B} T > \varepsilon_F$.  We use an effective dielectric constant of $\kappa_\mathrm{eff}=3.5$ appropriate for TBLG on h-BN.\label{fig:rhovsT&n}}
\end{figure*}

\begin{figure*}[t]
\begin{centering}
\includegraphics[width=\columnwidth]{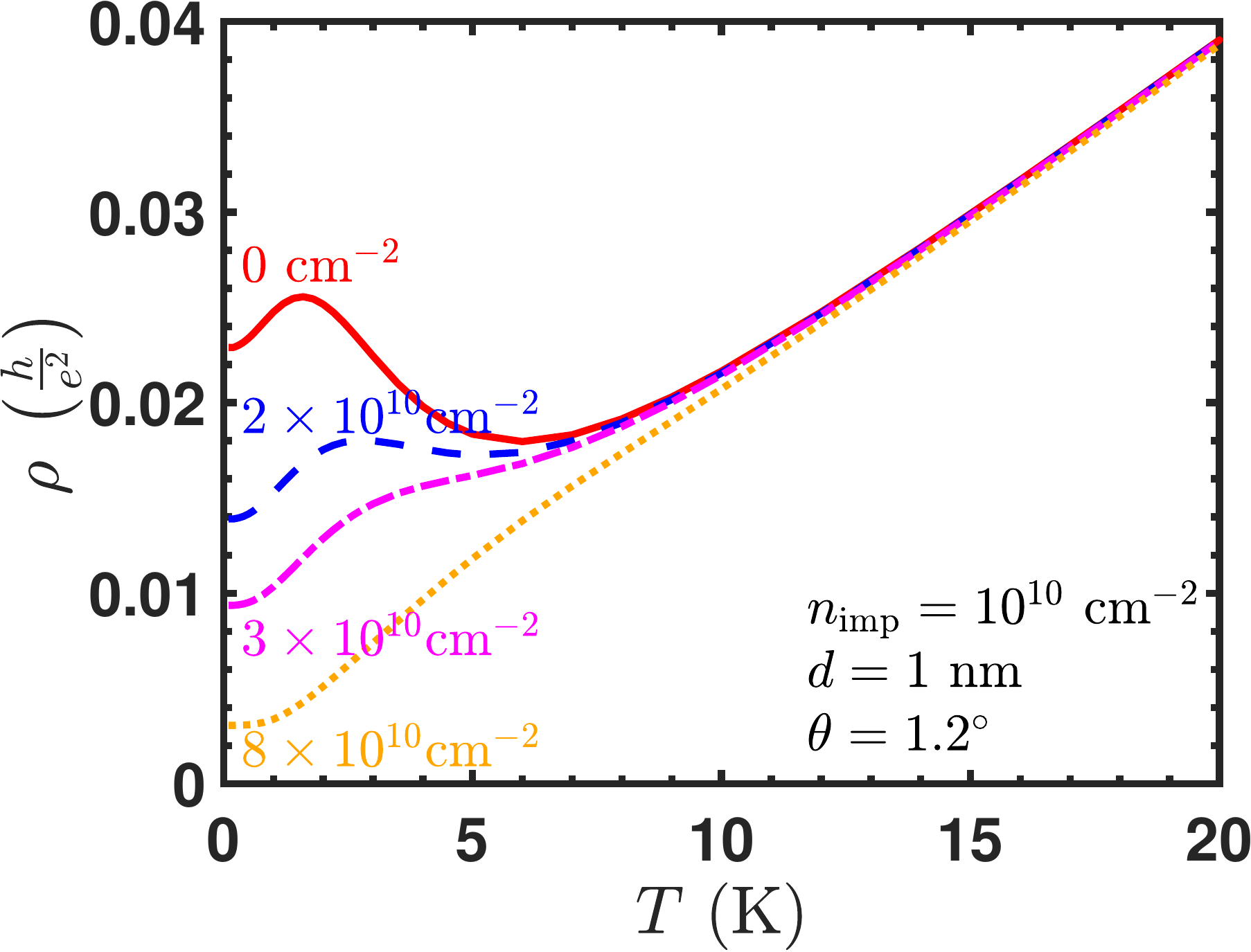}
\llap{\parbox[b]{6.8in}{\large{(a)}\\\rule{0ex}{2.35in}}}
\includegraphics[width=\columnwidth]{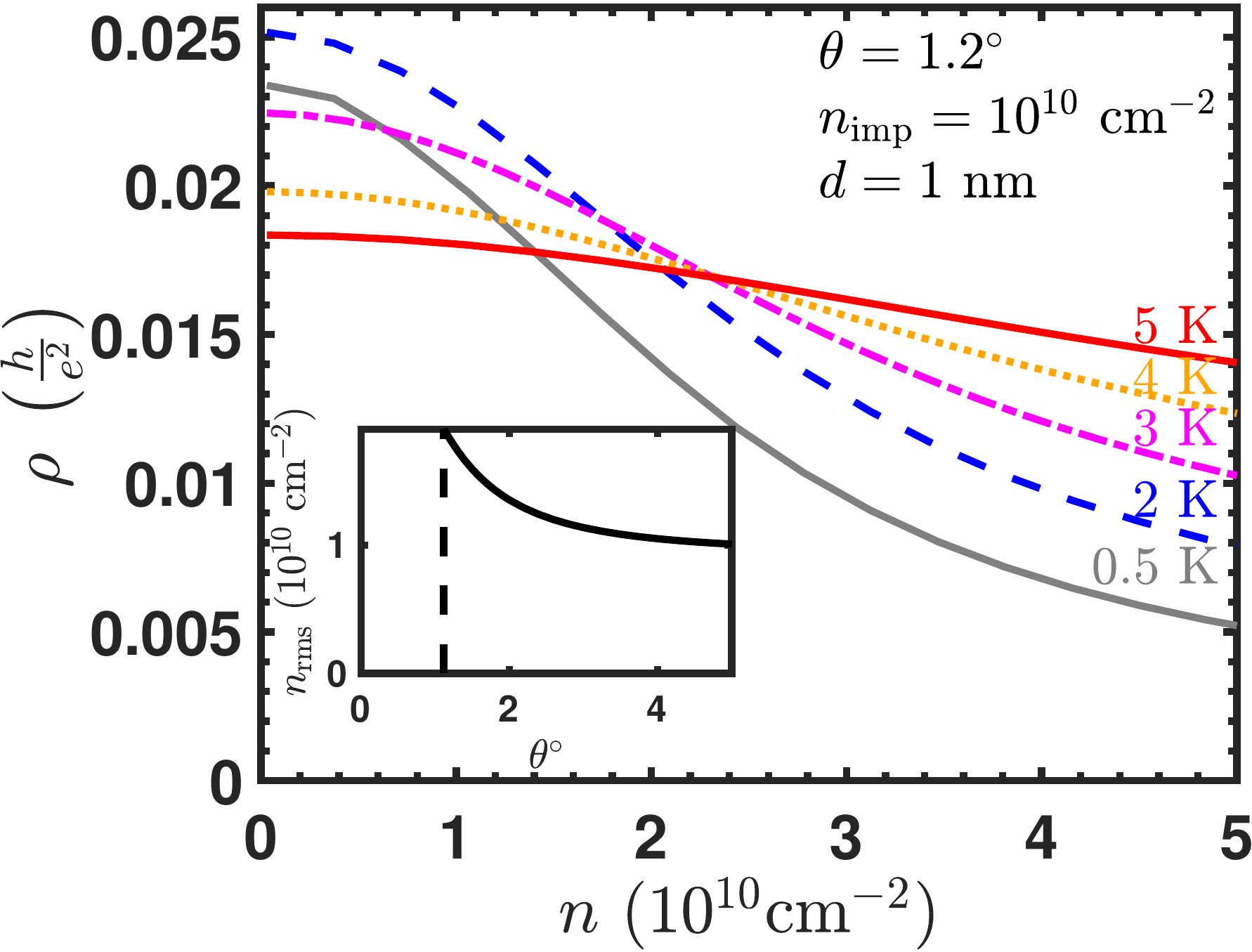}
\llap{\parbox[b]{6.8in}{\large{(b)}\\\rule{0ex}{2.35in}}}
\par\end{centering}
\caption{(Color online) Full effective medium theory for carrier transport in twisted bilayer graphene as a function of temperature (a) and density (b) considering both charged impurity and gauge phonon scattering.  At high temperature or high density, the gauge phonons dominate and there is a linear in T resistivity with negligible density dependence.  At low temperature and close to charge neutrality, there is a non-monotonic temperature dependence arising from charged impurity scattering. There is an inversion in temperature dependence as we move away from charge neutrality to higher densities because of a crossover from charged impurity to gauge phonon scattering. Inset: Charge density fluctuations vs twist angle.  While charged impurities are strongly screened by the large density of states, the electron-hole puddles are weakly affected by twist angle. \label{fig:rhovsn}}
\end{figure*}

The Boltzmann transport theory for charged impurity scattering is now well-established, and we refer the reader to Ref.~\cite{sarma2011electronic}  for details.  The scattering time is given by 
\begin{equation}
    \frac{1}{\tau_\mathrm{e-imp}(\varepsilon)} = \frac{ n_i |\varepsilon|}{2 \pi v^{*2}} \int_0^\pi d\theta \left|\frac{V_{ei}(q)}{\epsilon(q)}\right|^{2} (1-\cos^2\theta),
    \label{imptau}
\end{equation}
where $n_i$ is the impurity concentration and $V_{ei}(q)$ is the Coulomb impurity matrix element.  The $v^*$ dependent prefactor outside the integral comes from the enhanced $N_D$, while the dielectric function $\epsilon (q)$ is also enhanced by $v^{*-2}$.  As a result $1/\tau_{e-imp}$ scales roughly as $v^*$ and is suppressed near magic angle.  Throughout this work we use the RPA dielectric function $\epsilon(q,T)=1-V_{q}(q)\Pi_\mathrm{RPA}(q,T)$, where $V_{q}(q)=2\pi e^2/(\kappa q)$ is the Coulomb potential, $\kappa$ is the background dielectric constant and $\Pi_\mathrm{RPA}(q,T)$ is static RPA polarizability \cite{hwang_dielectric_2007}. The resistivity ($\rho_\mathrm{e-imp}$) is obtained from $\tau_\mathrm{e-imp}$ by the usual energy average
\begin{equation}
  \frac{1}{\rho_{\rm e-imp}}=e^{2}\int d\varepsilon\,N_{D}(\varepsilon)\frac{v^{*2}}{2}\tau_{\mathrm{e-imp}}(\varepsilon) \frac{-\partial n_{F}(\varepsilon-\mu)}{\partial\varepsilon}.
  \label{rhoimp}
\end{equation}

By contrast, the e-ph interaction in TBG is drastically different from that of monolayer graphene due to the emergence of hybrid folded phonon branches \cite{cocemasov_phonons_2013}. These hybrid phonons depend very sensitively on twist angle.  For the temperature range we consider, the lowest acoustic phonon branch (which is the same as monolayer graphene) dominates.  The gauge phonon contribution to the resistivity is \cite{Hwangacoustic2008,sohier_phonon-limited_2014}
\begin{equation}
\rho^\mathrm{e-ph}_{\beta_A}=\frac{16\Tilde{\beta}_{A}^{2}k_{F}}{e^{2}\mu_{s}v_{A}\left(v^{\star}\right)^{2}}F\left(\frac{T_{\mathrm{BG}}}{T}\right),
\label{rhophbeta}
\end{equation}
\noindent where $T_\mathrm{BG}$ is the Bloch-Gruneisen temperature ($k_B T_\mathrm{BG} = 2\hbar v_A k_F$), $\mu_s$ is the mass density of graphene, 
$k_F$ is the Fermi wave-vector and $v_A$ is the effective acoustic phonon velocity. Here $F(x)=\int_{0}^{1}dy[xy^{4}\sqrt{1-y^{2}}e^{xy}]/\left(e^{xy}-1\right)^{2}$.  The effective coupling constant $\Tilde{\beta}_A$ is proportional to strain, and therefore $\Tilde{\beta}_A \approx \beta_A (v^*/v_F)/[2\tan(\theta/2)]$ and $\beta_A = 3.6 \mathrm{eV}$ is the value obtained for monolayer graphene from DFPT and tight-binding calculations~\cite{lian_twisted_2018,sohier_phonon-limited_2014}.  The $D_A$ deformation potential phonons have a similar form to Eq.~\ref{rhophbeta}.  However, since they are heavily screened, this contribution becomes irrelevant in the temperature and density regimes we considered.

Figure \ref{fig:rhovsT&n} shows our results for the transport properties of tBG including both charged impurities and gauge phonons.  Gauge phonons dominate for $T>T_{\rm cross}$ and as $\theta \rightarrow \theta_M$.  The dependence of the crossover temperature and carrier density as a function of twist angle is shown in Fig.~\ref{diracvalid} and we observe that the phonon-dominated regime becomes more prominent as one approaches the magic angle.  For example, we find that the crossover temperature $T_\mathrm{cross}$ from impurity-dominated to phonon-dominated resistivity is $5$K at $n = 2 \times 10^{10} ~\mathrm{cm}^{-2}$ and twist angle $\theta = 1.2^o$.  We identify three distinct transport regimes: (i) For $T>T_{\rm cross}$, the resistivity is dominated by gauge phonons and is linear in temperature as one would expect from acoustic phonons; (ii)  For intermediate temperatures $k_B^{-1} \varepsilon_F < T < T_{\rm{cross}}$, the resistivity decreases with increasing temperature.  This is understood as the non-degenerate Fermi liquid scattering off charged impurities, where both the effects of screening and energy averaging reduce the resistivity as $\rho_{e-\mathrm{imp}}\propto(\varepsilon_{F}/k_{B}T)^{2}$; (iii) For $T < k_B^{-1} \varepsilon_{F}, T_{\rm cross}$, we have the degenerate Fermi liquid scattering theory of charged impurities where energy averaging has little effect on the resistivity and screening increases resistivity as $\delta \rho_{e-\mathrm{imp}} \propto  (k_{B}T/\varepsilon_{F})^2$.  This non-monotonic bump in the total resistivity $\rho(T)$ for $T \lesssim T_{\rm cross}$, and linear-in-T for $T > T_{\rm cross}$ are the main predictions of this work.

For a theory with homogeneous carrier density, the charged impurity limited resistivity diverges as $n\rightarrow 0$.  However, we know that these same charged impurities that dominate transport give rise to carrier density inhomogeneities that cure this divergence~\cite{adam_self-consistent_2007}.  In Fig.~\ref{fig:rhovsn} we show that the magnitude of the density inhomogeneity is weakly affected by the enhanced screening close to magic angle. We use the well-established effective medium theory~\cite{rossi_effective_2009} to average over these density inhomogeneities close to the Dirac point.  Our results for the resistivity as a function of temperature and carrier density for $\theta = 1.2^\circ$ are shown in Fig.~\ref{fig:rhovsn}.  There is a strong carrier density dependence at low temperature (dominated by charged impurities) and a weak density dependence at higher temperature (dominated by gauge phonons).  Our theory also predicts a temperature and density regime where resistivity decreases with increasing temperature causing an inversion in the temperature dependence at low and high density e.g. in Fig.~\ref{fig:rhovsn}b, close to charge neutrality, resistivity decreases with increasing temperature, while at $4\times10^{10}{\rm cm}^{-2}$ resistivity increases with increasing temperature.  (The curve for $T=0.5$K is strongly influenced by the carrier density inhomogeniety and doesn't follow this general trend). 

Finally, in our analysis we considered other possible relevant mechanisms for transport in tBG and found them to be negligible.  For example, Umklapp scattering~\cite{wallbank2019excess} is irrelevant for the densities we consider because only a small area of the moir\'e Brillouin zone is occupied and the hydrodynamic electron-hole scattering~\cite{Hohydro2018} contribution to resistivity remains weaker than impurity scattering for $T<100~{\rm K}$.  We conclude that for the experimentally relevant temperatures and densities the resistivity is determined only by the interplay between charged impurities and gauge phonons.   

\textit{Acknowledgement:} We acknowledge the Singapore Ministry of Education AcRF Tier 2 grant MOE2017-T2-2-140, the National University of Singapore Young Investigator Award (R-607-000-094-133) and use of the dedicated research computing resources at CA2DM.

%

\end{document}